\documentclass[epj,a4paper,english]{svjour}
\usepackage[T1]{fontenc}
\usepackage[latin1]{inputenc}
\usepackage{float}
\usepackage{graphicx}

\makeatletter
\usepackage{babel}
\makeatother

\begin{document}

\title{Simulation Study of $\chi_{c}\rightarrow J/\Psi+\gamma$ Detection
with $J/\Psi\rightarrow e^{+}e^{-}$ in pp Collisions
in the ALICE Experiment at LHC}


\author{P. Gonz\'alez\inst{1} \and P. Ladr\'on de Guevara\inst{1} \and E. L\'opez Torres\inst{3} \and A. Mar\'{\i}n\inst{2} \and E. Serradilla\inst{1} (for the ALICE Collaboration)}

%
%

\institute{CIEMAT, Department of Technology,\\ Madrid, Spain \and Gesellschaft f\"ur Schwerionenforschung mbH\\ Darmstadt, Germany \and CEADEN,\\Ciudad Habana, Cuba}

\date{Received: date / Revised version: date}
%
\abstract{
We present MonteCarlo preliminary results about the feasibility to
detect the $\chi_{c}$ family in p-p collisions at 14 TeV in the ALICE
Central Barrel at CERN LHC. The $\chi_{c1}$ and $\chi_{c2}$ were forced
to decay in the channel $J/\Psi + \gamma \rightarrow e^{+}e^{-} + \gamma$
and were merged with a proton-proton non-biased collision. After MonteCarlo
transport and simulation of the detector response, the $e^{+}$, $e^{-}$
and converted $\gamma$ were reconstructed and identified in the ALICE
ITS, TPC and TRD detectors. Separate signals corresponding
to $\gamma$ from $\chi_{c1}$ and from $\chi_{c2}$ were observed.
The position and relative weight of the fit to gaussians agreed with
the input values within the statistical limits. Similar studies will
be done for Pb-Pb collisions.
} 
%
\titlerunning{Simulation Study of $\chi_{c} \rightarrow J/\Psi + \gamma$ Detection with $J/\Psi \rightarrow e^{+} e^{-}$...}
\maketitle
\section{Introduction}
\label{intro}

\begin{figure}[ht]
\includegraphics[scale=0.40]{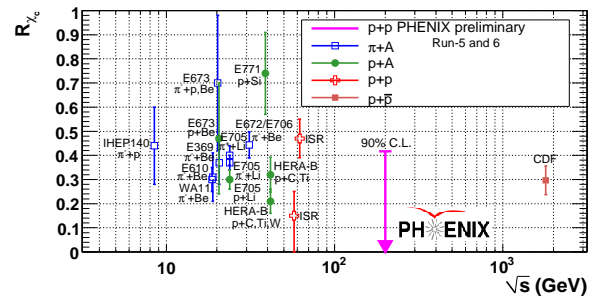} 
\caption{\label{fig:Rchic}Compilation of the experimental data on $R_{\chi_{c}}$ \cite{SUSU08}.}
\end{figure}

Heavy-flavour bound states constitute a valuable probe of the hot/dense
strongly interacting matter formed in relativistic collisions of heavy
nuclei. $J/\Psi$ suppression in central heavy ion collisions was
observed at SPS \cite{NA60} and RHIC \cite{RHIC01} energies. $\chi_{c}$
is an important source of $J/\Psi$ as a signal of deconfinement for
$R_{pA}$ and $R_{AA}$ studies, and p-p collisions should provide
the necessary baseline. The experimental study of $\chi_{c}$ family
has been done in the last 10 years in HERA \cite{HERAB}, CDF \cite{CDF01}
and L3 \cite{LEP01} experiments among others (see Fig. \ref{fig:Rchic}).
Recently, preliminary measurements of $R_{\chi_{c}}$ given by equation (\ref{eq:Rchic})
have been presented (PHENIX) at RHIC energies \cite{RHIC01}.
Predictions of $\chi_{c1}/\Psi$
and $\chi_{c2}$/$\Psi$ for Pb-Pb at LHC exist in the context of the
Statistical Hadronization Model \cite{LHC01}.
\begin{equation}
R_{\chi_{c}}=\frac{1}{\sigma(J/\Psi)}\sum_{i=1}^{2}\sigma(\chi_{ci})BR(\chi_{ci}\rightarrow J/\Psi+\gamma)\label{eq:Rchic}
\end{equation}

\section{ALICE Experiment at CERN LHC}
\label{sec:ALICE}

ALICE is a general purpose heavy ion experiment designed to study
the physics of strongly interacting matter and the Quark Gluon Plasma
(QGP) in nucleus-nucleus collisions at the LHC.

\begin{figure}[H]
\noindent \begin{centering}
\includegraphics[scale=0.4]{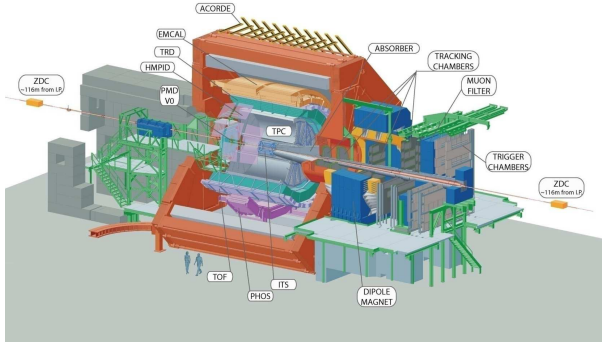} 
\par\end{centering}
\caption{\label{fig:alice-layout} General layout of the ALICE experiment
at CERN LHC.}
\end{figure}

The following ALICE subsystems were used for the identification and reconstruction
of the $\chi_{c}$ (see Fig. \ref{fig:alice-layout}):

\begin{itemize}
\item ITS (Inner Tracking System): measures the position of the primary
and secondary vertices of short lived particles.
\item TPC (Time Projection Chamber): performs the tracking of the charged
particles and the particle identification through dE/dx.
\item TRD (Transition Radiation Detector): allows $e/\pi$ separation and
improves momentum resolution.
\end{itemize}
Momentum for electrons from primary vertex was measured with the three
detectors. For some $\gamma$'s, part of the ITS or the complete ITS maybe
missing depending where the conversion occurs.

\section{Monte Carlo Generation}

The $\chi_{c}\rightarrow J/\Psi+\gamma$ events were generated with
AliRoot v4-11-Rev-02 under the following conditions:

\begin{itemize}
\item Cross sections \cite{HP03}:

\begin{equation}
 \sigma(\chi_{c1})=31.8\,\mu\textrm{b}\label{eq:x-chic1}
\end{equation}
\begin{equation}
 \sigma(\chi_{c2})=52.5\,\mu\textrm{b}\label{eq:x-chic2}
\end{equation}


\item Transverse momentum $p_{t}$ and rapidity $y$ for $\chi_{c}$ were
assumed to be the same as for $J/\Psi$, and given by equations (\ref{eq:pt})
and (\ref{eq:Y}), respectively.

\begin{equation}
f(p_{t})=\frac{p_{t}}{\left(1+\left(\frac{p_{t}}{4}\right)^{2}\right)^{3.6}}\label{eq:pt}\end{equation}

\begin{equation}
f(y)=\left\{ \begin{array}{cc}
e^{-(|y|-4)^{2}/2} & \textrm{ if }|y|>4\\
1 & \textrm{ if }|y|<4\end{array}\right.\label{eq:Y}\end{equation}

\item $J/\Psi$ were forced to $e^{+}e^{-}$ decay and the pseudorapidity
$\eta$ for the decay products ($e^{+}$, $e^{-}$ and $\gamma$ )
was constrained to $|\eta|<1.2$. The $\chi_{c}$ acceptance implied
by this condition is $\sim$10\%.
\item Each $\chi_{c}$ event was merged into a non-biased proton-proton collision at 14 TeV
generated by PYTHIA v6.214.
\end{itemize}
The transport of the interaction products through the detector and
its response were simulated by GEANT v1-9, and the data produced were
reconstructed using the AliRoot software.



\begin{figure}[ht]
\noindent \begin{centering}
\includegraphics[scale=0.26]{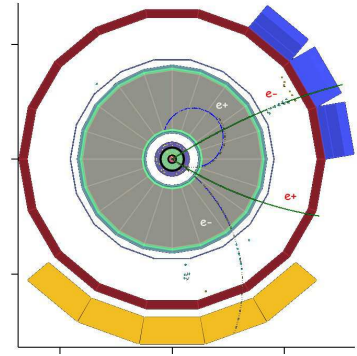} 
\par\end{centering}
\caption{\label{fig:vis} Visualization of a $\chi_{c} \rightarrow J/\Psi + \gamma$ 
event in the ALICE Central Barrel.}
\end{figure}

$10^{6}$ of these events were generated and reconstructed using the
ALICE GRID. The reconstruction of one of these events
is shown in Fig. \ref{fig:vis}, where non-electron tracks, mainly
generated in the proton-proton non-biased collisions, have
been removed for clarity.

%
%

\section{$J/\Psi$ Reconstruction}

\begin{figure}[ht]
\includegraphics[scale=0.44]{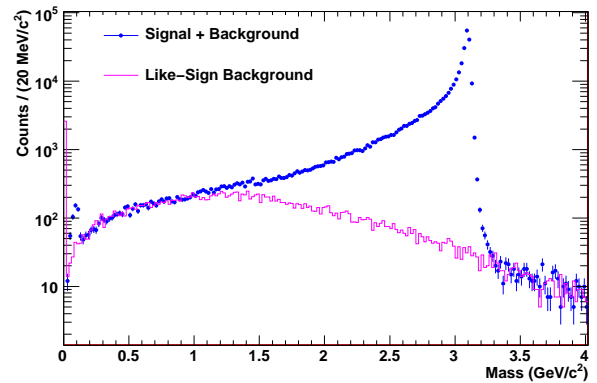} 
\caption{\label{fig:JPsi}Mass spectrum of $J/\Psi$ (circles) and its combinatorial
background (line).}
\end{figure}

After selecting tracks coming from the primary vertex within 3 sigmas,
and requiring a signal in the central detector ITS, TPC and TRD, electrons
were reconstructed and identified with an efficiency of 94.1\% and
purity of 99.5\%. Then, the invariant mass $M(e^{+}e^{-})$ spectrum
was computed. Contamination of $\sim$
4 \% coming from $\gamma$ converted very near of the origin, from
Dalitz pairs in $\pi^{0}$ decay and from other sources was cured
cutting out the pairs where angle between $e^{+}$ and $e^{-}$
was lower than 0.05 rad.
Final spectrum (see Fig. \ref{fig:JPsi}) showed the $J/\Psi$. Note 
that the tail on the left of the $J/\Psi$
peak is due to bremsstrahlung. The like-sign technique reproduced
quite well the combinatorial background
(see Fig. \ref{fig:JPsi}) since open heavy flavour semileptonic decay was
not taken into account in the generated events.

The reconstruction efficiency for $J/\Psi$ within $|\eta|<0.9$ is
37.0\% after subtracting the combinatorial background and integrating
the peak from 2.8 to 3.6 GeV/$c^{2}$ to suppress the bremsstrahlung
tail. Including the bremsstrahlung tail the efficiency is 47.6\%.



\section{$\gamma$ Reconstruction}

\begin{figure}[H]
\includegraphics[scale=0.44]{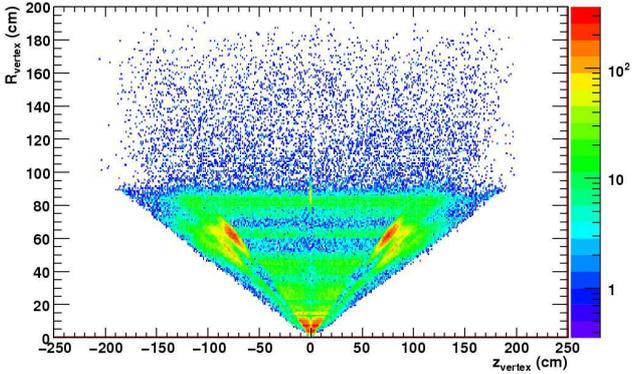} 
\caption{{\footnotesize \label{fig:gamma-vertex} Distribution of reconstructed
$\gamma$ conversion vertices.}}
\end{figure}

In our data, photons originate from the decay of $\chi_{c}\rightarrow J/\Psi+\gamma$
and from the decay of neutral mesons from the simulated non-biased
p-p collisions. Part of these photons convert on the detector material,
mainly in the material prior to the TPC (see Fig. \ref{fig:gamma-vertex}),
which allows a good lever arm to reconstruct the by products. The
conversion probability in the $\chi_{c}$ acceptance is of 8.3\%.

The reconstruction of the photons from the positive and negative tracks
was done by the {}``conversion method'', based on the finding of
the opposite sign tracks associated to a $V^{0}$ and on cuts on the
angle ($<$ 0.1 rad) and on the mass ($<$ 0.175 GeV/$c^{2}$) of the positive
and negative tracks.



\begin{figure}[H]
\includegraphics[scale=0.44]{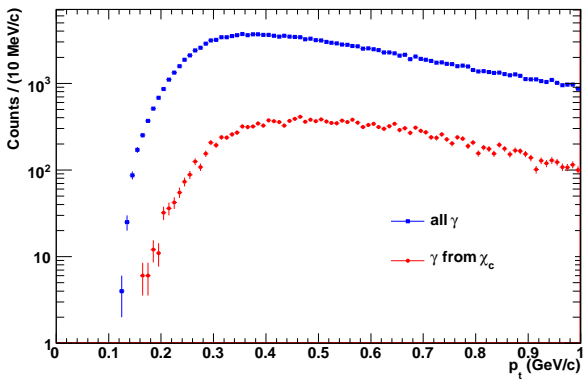}
\caption{{\footnotesize \label{fig:gamma-pt} Transverse momentum distribution,
$p_{t}$, of all $\gamma$ and of $\chi_{c}$ photons, reconstructed
using the conversion method.}}
\end{figure}

The Fig. \ref{fig:gamma-pt} shows the $p_{t}$ for all the reconstructed
$\gamma$ (upper), and the $p_{t}$ for the $\gamma$ from $\chi_{c}$
confirmed by MonteCarlo (lower). Note that $\gamma$ can be reconstructed
down to a $p_{t}$ of 100 MeV/c.


\begin{figure}[H]
\includegraphics[scale=0.44]{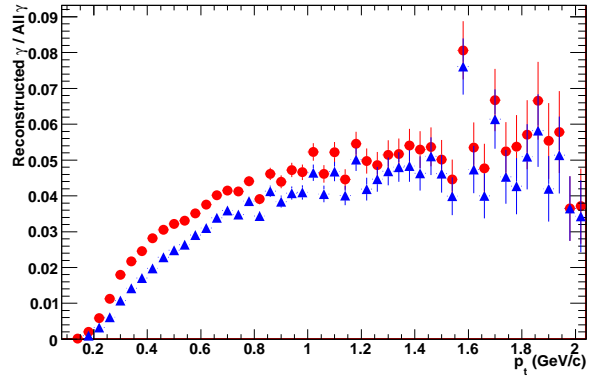}
\caption{\label{fig:gamma-efficiency} Reconstruction efficiency of $\chi_{c}$ photons. 
Circles are the efficiency, given the conversion probability and the $V^0$ method, using the MC 
information to identify the $\chi_{c}$ photons, and triangles
the efficiency from reconstruction after applying the cuts and selecting only the photons coming from $\chi_{c}$.}
\end{figure}

The reconstruction efficiency of $\chi_{c}$ photons has been
computed for two sets (see Fig. \ref{fig:gamma-efficiency}):
a) the reconstructed $V^0$ tagged as $\gamma$ after applying the selection
cuts and requested to come only from $\chi_{c}$, and 
b) the reconstructed $V^{0}$ associated to $\chi_{c}$ photons.
Figure \ref{fig:gamma-efficiency} shows an average total efficiency for $\gamma$ from $\chi_{c}$
decay (that lie in the low $p_{t}$ region) of 3.0\%.
Work is in progress to improve efficiency. 
It also shows that the set of cuts selects quite well the $\gamma$ from $\chi_{c}$.




\section{ $\chi_{c}$ Reconstruction}

\begin{figure}[H]
\includegraphics[scale=0.44]{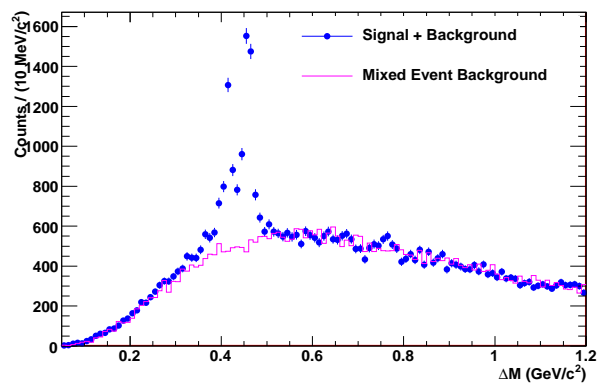}
\caption{{\footnotesize \label{fig:chic-dm-bg} $\Delta M=M(e^{+}e^{-}\gamma)-M(e^{+}e^{-})$
spectrum (circles) and its combinatorial background (line).}}
\end{figure}

Once the $J/\Psi$ and the $\gamma$'s coming from the primary vertex
are detected on the central barrel,
$\chi_{c}$ can be identified in the invariant mass spectrum of $J/\Psi$
and $\gamma$, $M(e^{+}e^{-}\gamma)$. However, the invariant mass
difference $\Delta M=M(e^{+}e^{-}\gamma)-M(e^{+}e^{-})$ provides
better resolution than $M(e^{+}e^{-}\gamma)$ because of the cancellation
of systematic errors. Some $\gamma$'s come in fact from
electron bremsstrahlung conversions near the primary vertex. To cure this
contamination we requested the angle between the electron and the 
reconstructed $\gamma$ to be greater than 0.05 rad.
The combinatorial background was computed with the event mixing
technique (see Fig. \ref{fig:chic-dm-bg}).

\begin{figure}[H]
\includegraphics[scale=0.44]{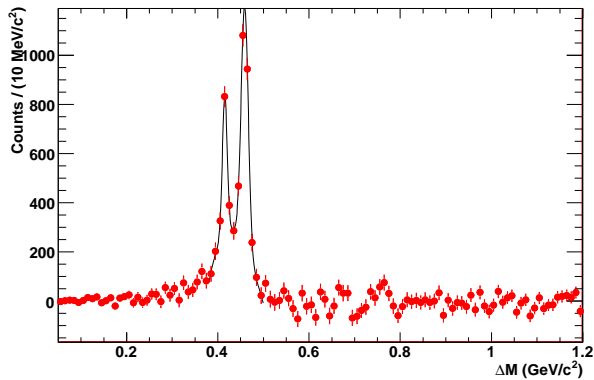} 
\caption{{\footnotesize \label{fig:chic-signal}$\Delta M$ after background
subtraction, showing energy transitions from $\chi_{c1}$ and $\chi_{c2}$
to $J/\Psi$ in CM reference system. The continuous line shows the nominal position
and amplitude of the peaks.}}
\end{figure}

Fig. \ref{fig:chic-signal} shows the spectrum after subtraction of the background.
Integration from 0.3 to 0.5 GeV/c allows to compute the mean reconstruction
efficiency within the Central Barrel acceptance for $\chi_{c}$ as
0.9\%.

\begin{figure}[H]
\includegraphics[scale=0.44]{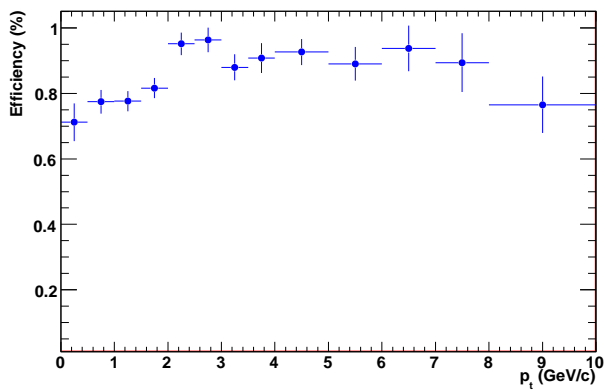} 
\caption{{\footnotesize \label{fig:chic-eff}$\chi_{c}$ reconstruction efficiency
as a function of $p_{t}$.}}
\end{figure}

$\chi_{c}$ reconstruction efficiency for different intervals of $p_{t}$
ranges from 0.7\% to 1.0\% within the small statistics of this study
(see Fig. \ref{fig:chic-eff}).

\section{ $\chi_{c}$ Expected Rates}

Taking into account the cross-sections given in 
(\ref{eq:x-chic1}) and (\ref{eq:x-chic2}), the decay
rates \cite{PDG}:

\begin{equation}
 \chi_{c1} \rightarrow J/\Psi + \gamma : 35.6 \% 
\end{equation}
\begin{equation}
 \chi_{c2}\rightarrow J/\Psi + \gamma : 20.2 \% 
\end{equation}
\begin{equation}
 J/\Psi \rightarrow e^{+} e^{-} :  5.94 \%
\end{equation}

\noindent and assuming a luminosity:

\begin{equation}
L = 10^{30} \textrm{cm}^{-2} \textrm{s}^{-1}
\end{equation}

\noindent the rate of $\chi_{c}\rightarrow J/\Psi + \gamma$ production is 1.3 per second.
This must be weighted by our acceptance (10 \%) and our
reconstruction efficiency of 0.9 \% to give $1.2 \times 10^{-3}$ observable
$\chi_{c}$ per second, assuming a 100 \% trigger efficiency. The trigger strategy
is in discussion, so this figure is a maximum, and assuming a
nominal pp run of $10^7$ s, the total maximum expected $\chi_{c}$ is accordingly 
$\sim$ 12000.

\section{Conclusions}

The electrons from $J/\Psi$ were reconstructed and identified using the
ALICE Central Barrel with an efficiency of 94.1\% and purity of 99.5\%.
The $J/\Psi$ was reconstructed with an efficiency of $\sim$37\%
after cutting the bremsstrahlung tail, selecting only $M(e^{+}e^{-})$
events from 2.8 GeV/$c^{2}$ to 3.6 GeV/$c^{2}$.


$\gamma$ were reconstructed via conversions in the Central Barrel
with $p_{t}$ reaching down to 100 MeV/c. The mean reconstruction
efficiency was $\sim$3\%.

$\chi_{c}$ can be detected in the ALICE Central Barrel down to a
$p_{t}$ of 0.250 GeV/c. ALICE TPC resolution allows for observing a
defined structure of $\chi_{c}$ family via $M(e^{+}e^{-}\gamma)-M(e^{+}e^{-})$
mass difference, showing transitions from $\chi_{c1}$ and $\chi_{c2}$
to $J/\Psi$. The overall efficiency of $\chi_{c}$ reconstruction
was $\sim$0.9\%.

Assuming a luminosity of 10$^{30}$ cm$^{-2}$ s$^{-1}$ the rate of
observable $\chi_{c}$'s is expected to be $1.2 \times 10^{-3}$ per second,
with an ideal trigger.

%

\begin{thebibliography}{}
%
%

\bibitem{NA60}  \textit{J/$\Psi$ production in In-In and pA collisions.} E. Scomparin (for the NA60 Collaboration);  J. Phys. G.:Nucl. Part.Phys  \textbf{34}(2007) (S463-S469) and references therein.


\bibitem{RHIC01} \textit{RHIC results in J/$\Psi$} M. J. Leitch; J. Phys. G.: Nucl. Part.Phys \textbf{34}(2007) (S453-S462) and references therein.


\bibitem{HERAB} \textit{J/$\Psi$ Production via $\chi_{c}$ Decays in 920 GeV pA Interactions.} I. Abt {\it et al.} (HERA-B Collaboration).
 Phys. Lett.  \textbf{B 561}(2003) 61-72.


\bibitem{CDF01} \textit{ Production of J/$\Psi$ Mesons from $\chi_{c}$ Meson Decay in pp Collisions at $\sqrt{s}$ = 1.8 TeV.} F. Abe {\it et al.}(CDF Collaboration). Phys. Rev. Lett.  \textbf{79},(1977) 578-583.


\bibitem{LEP01} \textit{$\chi_{c}$ Formation in Two-Photon Collisions at LEP. The L3 Collaboration.} CERN-EP/98-184

\bibitem{LHC01} \textit{Heavy Ion Collisions at the LHC-Last Call for Predictions.} S. Abreu {\it et al.}
  J.\ Phys.\ G \textbf{35}, (2008) 054001.
  [arXiv:0711.0974 [hep-ph]].

\bibitem{SUSU08} \textit{J/$\Psi$ production in Cu+Cu and Au+Au collisions at RHIC-PHENIX.} Susumu X. Oda. ArXiv: 0804.4446v2[nucl-ex] \textbf{29} Apr 2008.

\bibitem{HP03} \textit{Hard Probes In Heavy Ion Collisions At The LHC:Heavy Flavour Physics.} ArXiv: hep-ph/0311048 \textbf{v1.4}, Nov 2003.

\bibitem{PDG} \textit{The Review of Particle Physics.} C. Asler {\it et al.}, Physics Letters \textbf{B667}, (2008) 1.


\end{thebibliography}
%

\end{document}